%% file: main.tex
\documentclass[conference]{IEEEtran}
\IEEEoverridecommandlockouts
\usepackage{cite}
\usepackage{amsmath,amssymb,amsfonts}
\usepackage{algorithmic}
\usepackage{graphicx}
\usepackage{textcomp}
\usepackage{xcolor}
\def\BibTeX{{\rm B\kern-.05em{\sc i\kern-.025em b}\kern-.08em
    T\kern-.1667em\lower.7ex\hbox{E}\kern-.125emX}}
\begin{document}

\title{Rallying Adversarial Techniques against Deep Learning for Network Security}


\author{\IEEEauthorblockN{Joseph Clements*, Yuzhe Yang$^\dag$, Ankur A. Sharma*, Hongxin Hu$^{\dag\dag}$, Yingjie Lao*}
\IEEEauthorblockA{
{*Deptartment of Electrical and Computer Engineering, Clemson University}\\
{$^{\dag}$School of Computing, Clemson University}\\
{Clemson, South Carolina}\\
\{jfcleme, ankurs, yuzhey\}@g.clemson.edu, ylao@clemson.edu\\
{Department of Computer Science and Engineering}\\
{University at Buffalo}\\
hongxinh@buffalo.edu\\
}
}

\author{\IEEEauthorblockN{Joseph Clements, Yuzhe Yang, Ankur A. Sharma, Hongxin Hu$^{\dag}$, Yingjie Lao}
\IEEEauthorblockA{
{Clemson University, Clemson, South Carolina}\\
\{jfcleme, yuzhey, ankurs\}@g.clemson.edu, ylao@clemson.edu\\
{$^{\dag}$University at Buffalo, Buffalo, New York}\\
hongxinh@buffalo.edu\\
}
}


\maketitle

\begin{abstract}
\input{tex/1-abstract.tex}

\end{abstract}

\begin{IEEEkeywords}
Network Intrusion Detection System, Adversarial Machine Learning, Adversarial Examples, Deep Learning.
\end{IEEEkeywords}

\input{tex/2-introduction.tex}

\input{tex/3-background.tex}

\input{tex/4-NIDS.tex}

\input{tex/5-setup.tex}

\input{tex/6-networksecurity.tex}

\input{tex/7-adversarialML.tex}

\input{tex/8-optimize.tex}
\input{tex/999-conclusions.tex}

\input{tex/999-acknowledgements.tex}

\bibliographystyle{IEEEtran}
\bibliography{tex/references}

\end{document}

%% file: tex/1-abstract.tex
Recent advances in artificial intelligence and the increasing need for robust defensive measures in network security have led to the adoption of deep learning approaches for network intrusion detection systems (NIDS). These methods have achieved superior performance against conventional network attacks, enabling unique and dynamic security systems in real-world applications. Adversarial machine learning, unfortunately, has recently shown that deep learning models are inherently vulnerable to adversarial modifications on their input data. In this work, we explore the potential of adversarial entities to compromise such vulnerabilities to compromise deep learning-based NIDS systems. Specifically, we show that by modifying on average as little as $1.38$ of an observed packet's input features, an adversary can generate malicious inputs that effectively fool a target deep learning-based NIDS. Therefore, it is crucial to consider the performance from the conventional network security perspective and the adversarial machine learning domain when designing such systems.

%% file: tex/2-introduction.tex
\section{Introduction} \label{sec:introduction}
Mainly attributable to advances in deep learning, the field of artificial intelligence has been growing swiftly in the recent past. Through many examples, it has been witnessed that deep learning systems have the potential to achieve or even surpass human-level performance on specific tasks. Furthermore, these systems are not explicitly given a function to implement but instead can discover hidden rules or patterns that developers may not comprehend. This ability to learn has made deep learning an indispensable tool for advancing the state-of-the-art in multiple fields.

With these remarkable successes, it is unsurprising that deep learning techniques are quickly being adopted in network security for use in intrusion detection~\cite{buczak2016survey}, malware analysis~\cite{gardiner2016security}, spam filtering~\cite{blanzieri2008survey}, and phishing detection ~\cite{abu2007comparison}. However, the growing popularity of novel network paradigms (i.e., Internet of Things (IoT) and mobile networks) also brings unique and challenging security requirements. To this end, modern deep learning algorithms can rival traditional approaches, especially in these emerging fields. Recently, the field of deep learning-based network intrusion detection systems (DL-NIDS) has been growing due to the variability and efficiency of the deep learning model. The availability of novel techniques such as recurrent neural networks, semi-supervised learning, and reinforcement learning is allowing DL-NIDS to achieve success in applications that have been traditionally out of the reach of intrusion detection systems ~\cite{yin2017deep,ashfaq2017fuzziness,venkatesan2017detecting}.

However, the downside of deep learning is that the high non-linearity seen in these systems limits the ability of developers to guarantee or explain their functionality. This complexity allows for the possibility of unseen security risks. Indeed, many recent works have demonstrated the vulnerability of deep learning to adversarial manipulation~\cite{li2018hu, liu2017neural, szegedy2013intriguing}. For example, adversarial examples can completely misclassify a deep learning model by only slightly altering the network input data~\cite{carlini2017towards, goodfellow2014explaining, papernot2016limitations, chen2018ead}. In response to the threat that this form of attack poses to deep learning, multiple potential defenses have arisen~\cite{ papernot2016distillation, kurakin2016adversarial, song2017pixeldefend}. Despite this, security applications remain vulnerable since it is uncertain which defensive methodologies are most effective in given scenarios.

Therefore, to ensure the defensive capabilities of deep learning-based security systems, these applications should be evaluated against the traditional performance metrics in the target security field and those vulnerabilities from the adversarial deep learning domain. If deployed without understanding these vulnerabilities, a deep learning model could quickly become the most sensitive component of a security system. This paper analyzes adversarial example attacks against current deep learning-based network intrusion detection system (DL-NIDS), demonstrating the real-world vulnerability of such systems. Specifically, we investigate the security of a DL-NIDS that has been used for security analysis, Kitsune. This system offers a similar level of defensive capability as traditional intrusion detection systems while requiring a lower overhead~\cite{mirsky2018kitsune}. We evaluate the DL-NIDS from two perspectives: 1) the ability to defend from malicious network attacks and 2) the robustness against adversarial examples. Our experimental evaluations find that this model is vulnerable to adversarial manipulation through its deep learning model. Thus, we argue that the benefit of such systems should be evaluated by their ability to defend from traditional attacks and the vulnerability of the deep learning model to manipulation.

In the remainder of the paper, we first introduce the basics of deep learning and its use in intrusion detection systems and the state-of-the-art in deep adversarial learning. We decompose the target DL-NIDS, Kitsune, in Section \ref{sec:NIDS}. Then, we briefly outline our experimental setup in Section \ref{sec:setup}. In Sections \ref{sec:NS} and \ref{sec:AM}, we evaluate the DL-NIDS from the perspectives of network security and deep adversarial learning, respectively. Finally, Section \ref{sec:conclusions} concludes the paper.

%% file: tex/3-background.tex
\section{Background} \label{sec:background}

\subsection{Deep Learning based Network Intrusion Detection Systems}
\begin{figure}[htbp]
\resizebox{0.49\textwidth}{!}{%
\includegraphics[]{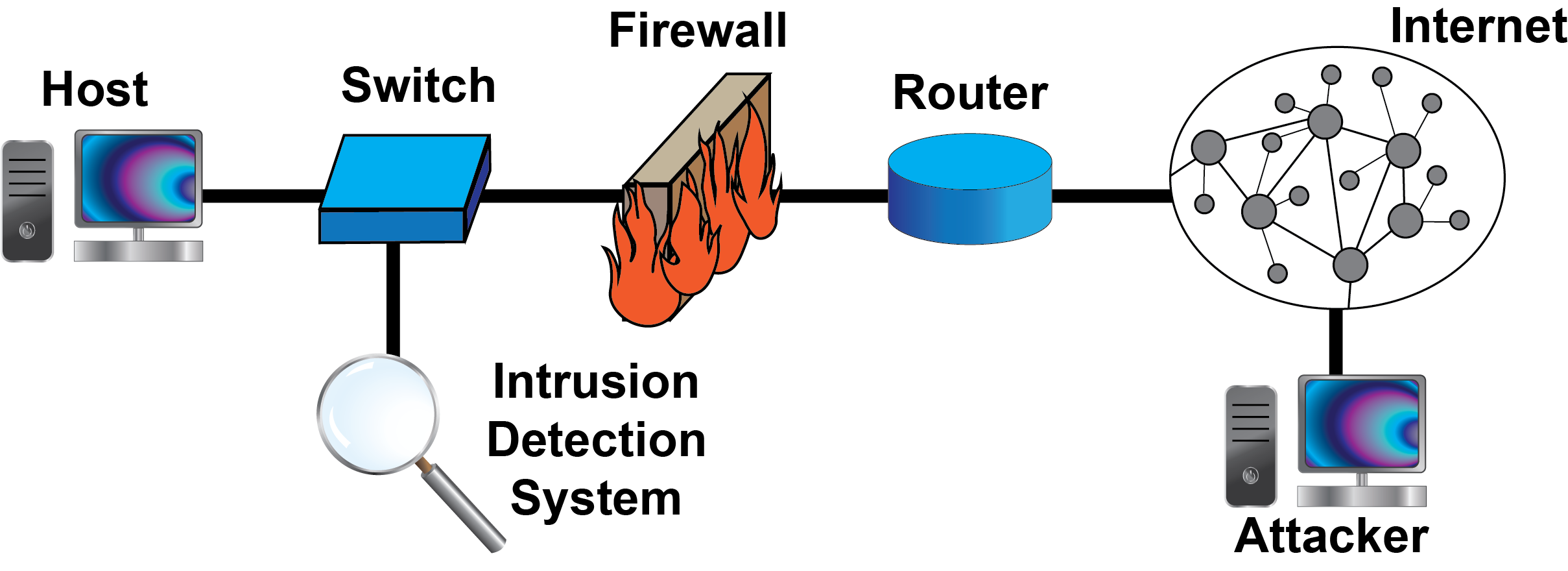}}
\centering
\caption{An intrusion detection system positioned to defend a host device from abnormal network traffic.}
\label{fig:ids}
\end{figure}

In recent years, the increasing frequency and size of cyber-attacks in recent years~\cite{kuypers2016empirical} have made network intrusion detection systems (NIDS) a critical component in network security. An example of a network intrusion detection system is shown in Figure \ref{fig:ids}. The intrusion detection system essentially acts as a gatekeeper at the target node, which activates a firewall or alerts a host device when malicious network traffic is detected.  Unfortunately, while these systems can effectively defend the entry point, much of the network remains unprotected. In other words, attacks that remain internal to the network are often difficult to detect by the traditional intrusion detection systems~\cite{mirsky2018kitsune}.

Deploying an intrusion detection system at multiple nodes distributed throughout the network can fill this hole to further secure networks. However, a significant drawback of the traditional rule-based approach is that each intrusion detection system must be explicitly programmed to follow a set of rules. This process also generates potentially long lists of rules that need to be stored locally to access intrusion detection systems. Furthermore, any changes in a network node might potentially lead to an update for the entire network. To this end, DL-NIDS have the potential to overcome this weakness as they can generalize the defense by capturing the distribution of typical network traffic instead of being explicitly programmed~\cite{mirsky2018kitsune, srivastav2013novel, damopoulos2012evaluation, li2012efficient}. In addition, these methods do not require large lookup tables, which could also reduce the implementation cost.

\subsection{Adversarial Example Generation}
A major focus of adversarial deep learning is the adversarial example generation, which attempts to find input samples by slightly perturbing the original benign data to yield different classifications. Formally, the adversarial example generation process can be expressed by~\cite{carlini2017towards}:
\begin{equation}\label{eq:adversarialExample}
  \begin{array}{lc}
  \mbox{minimize} & \mathcal{D}( \vec{x},\vec{x}+\vec{\delta} ) \\
  \mbox{such that} & \mathcal{C}( \vec{x}+\vec{\delta},\vec{t} ) \\
  & \vec{x}+\vec{\delta} \in \mathbb{X}
  \end{array}
\end{equation}
Where $\vec{x}$ is the model's original primary input, $\vec{\delta}$ is a perturbation on $\vec{x}$ to achieve the desired adversarial behavior, and $\mathbb{X}$ defines a bounded region of the valid input values. $\mathcal{D}(\cdot)$ is a distance metric that limits $\delta$, while $\mathcal{C}(\cdot)$ is a constraint that defines the goal of the attack. Two commonly used constraint functions are $F(\vec{x})=\vec{t}$ and $F(\vec{x}) \neq \vec{t}$. The first defines a targeted attack in which the adversarial goal is to force the network output, $F(\vec{x})$, to a specific output, $\vec{t}$. The second defines the untargeted scenario where the adversarial goal is for the network to produce any output except $\vec{t}$. The choice of $\mathcal{D}(\cdot)$ also greatly affects the outcome of the attack. In the existing works, $L_P$ norms (i.e., $L_0$, $L_1$, $L_2$, and $L_\infty$) are often used due to their mathematical significance and correlation with perceptual distance in image or video recognition. Recently, new distance metrics are being explored with the recent works such as spatially transformed adversarial examples~\cite{xiao2018spatially}.

Many algorithms for generating adversarial examples utilizing various $C(\cdot)$, $D(\cdot)$, and optimization approaches have been developed in the literature. For example, one of the earliest adversarial example algorithms, Fast Gradient Sign Method (FGSM), perturbs every element of the input in the direction of its gradient by a fixed size~\cite{goodfellow2014explaining}. While this method produced quick results, the Basic Iterative Method (BIM) can significantly decrease the perturbation, requiring a longer time to run~\cite{kurakin2016adversarial}. Furthermore, adversarial example generation algorithms continue to grow more sophisticated as novel attacks build on the foundation of existing works. An example of this is the elastic net method (ENM) which adds an $L_1$ regularization term and the iterative shrinkage-thresholding algorithm to Carlini and Wagner's attack~\cite{chen2018ead}. Moreover, adversarial examples are expanding out from image processing into alternate fields where they continue to inhibit the functionality of deep learning models~\cite{carlini2018audio, huang2017adversarial,kos2018adversarial}. The effort to draw researcher awareness to the subject has even lead to the generation of competitions in which contestants attempt to produce and defend neural networks from this adversarial example~\cite {brendel2018adversarial, kurakin2018adversarial}.

\subsection{Robustness against Adversarial Examples}
Some researchers believe that the vulnerability of deep learning models to adversarial examples is evidence of a pervasive lack of robustness rather than simply an inability to secure these models~\cite{ford2019adversarial, fawzi2016robustness, gilmer2018adversarial}. As such, defenses attempt to bolster the deep learning model's robustness by using either reactive or proactive methods~\cite{yuan2019adversarial}. Defensive distillation and adversarial training are two proactive defenses, which improve a neural network's robustness by retraining the network weights to smooth the classification space~\cite{papernot2016distillation, kurakin2016adversarial}. A recent example of a reactive defense is, PixelDefend, which attempts to perturb adversarial example input back to the region of inputs space that is correctly handled by the network~\cite{song2017pixeldefend}.

When deep learning is powering security applications, the robustness of the model is even more critical. The field of malware classification is a prime example as deep learning models have been shown to perform superbly in this area in multiple implementations and scenarios~\cite{zhu2016featuresmith, sayfullina2015efficient, saxe2015deep, arp2014drebin}. Unfortunately, when adversarial examples are presented to these systems, the lack of robustness in the deep learning model often allows an attacker to bypass these security measures~\cite {grosse2017adversarial, kolosnjaji2018adversarial}. Despite this vulnerability, deep learning is a prime candidate for security implementations when traditional defenses' resource demands or static nature inhibit their practicality. Thus, as deep learning continues to develop into network intrusion detection, the robustness of such systems should be thoroughly studied. To this end, researchers are continuing to develop guidelines and frameworks to aid in ensuring the robustness of machine learning systems against adversarial manipulations~\cite{gilmer2018motivating, carlini2019evaluating}.

%% file: tex/4-NIDS.tex
\begin{figure*}[htbp]
\resizebox{0.99\textwidth}{!}{
\includegraphics[]{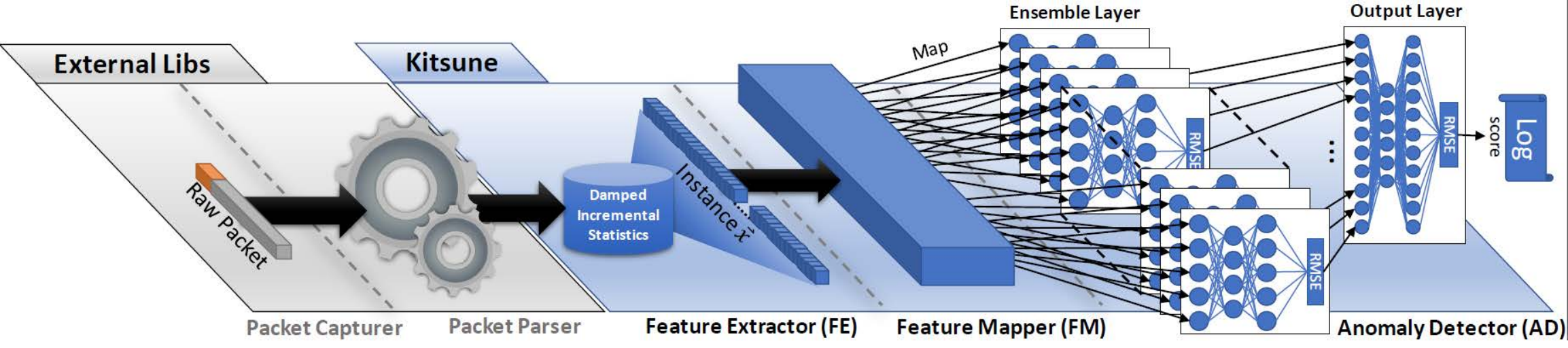}}
\centering
\caption{A graphical representation of Kitsune~\cite{mirsky2018kitsune}.}
\label{fig:kitsune}
\end{figure*}

\section{Evaluated Network} \label{sec:NIDS}

This section presents a brief overview of the network intrusion detection system and then analyzes Kitsune's deep learning model, KitNET, in more detail.

\subsection{Kitsune Overview}
The DL-NIDS, Kitsune, is composed of Packet Capturer, Packet Parser, Feature Extractor, Feature Mapper, and Anomaly Detector~\cite{mirsky2018kitsune}. The Packet Capturer and Packet Parser are standard components of NIDS, which forward the parsed packet and meta-information (e.g., transmission channel, network jitter, capture time) to the Feature Extractor. Then, the Feature Extractor generates a vector of over $100$ statistics which defines the packet and current state of the active channel. The Feature Mapper clusters these features into subsets fed into the Anomaly Detector, which houses the deep learning model, KitNET.

The Kitsune DL-NIDS is specifically targeted at being a lightweight intrusion detection system deployed on network switches in the IoT settings. Thus, each implementation of Kitsune should be tailored to the network node that it defends. This goal is achieved by using an unsupervised online learning approach that allows the DL-NIDS to dynamically update in response to the traffic at the target network node. The algorithm assumes that all real-time transmissions during the training stage are legitimate and thus learns a benign data distribution. For inference, it analyzes the incoming transmissions to determine if it resembles the learned distribution.

\subsection{KitNET}

KitNET, Kitsune's deep-learning backbone, consists of an ensemble layer and an output layer. The ensemble layer includes multiple autoencoders, each working on a single cluster of inputs provided from the Feature Mapper. The output scores of these autoencoders are normalized before being passed to an aggregate autoencoder in the output layer, whose score is used to assess the security of the network traffic data.

\subsubsection{The Autoencoders}

The fundamental building block of KitNET is an autoencoder, a neural network that reduces an input down to a base representation before reconstructing to the same input dimension from that representation. The autoencoders in KitNET are trained to capture the properties of typical network traffic correctly. The number of hidden neurons inside an autoencoder is limited so the network can learn a compact representation.

KitNET employs a root-mean-squared-error (RMSE) function on each autoencoder as the performance criteria. The score generated by each autoencoder block is given by:
\begin{equation}\label{eq:RMSE}
 s(\vec{x})= RMSE(\vec{x},F(\vec{x})) = \sqrt{\dfrac{\sum_{i=1}^n({x_i}-F({\vec{x}})_i)^2}{n}}
\end{equation}
where $n$ is the number of inputs. Because the model was trained to reproduce instances from $X$, a low score indicates the input resembles the normal distribution well.

\subsubsection{The Normalizers}
Another component used by Kitsune is the normalizers, appearing both before entering KitNET and before the aggregate autoencoder. These normalizers implement the standard function:
\begin{equation}\label{eq:norms}
  norm(x_i) = \dfrac{x_i-min_i}{Max_i-min_i}
\end{equation}
which linearly scales minimum and maximum input values to $0$ and $1$, respectively. In Kitsune's training, the value of $Max_i$ and $min_i$ respectively take on the maximum and minimum input values seen by the $x_{i_{th}}$ element during training.

\subsection{Classifying the Output}
The primary output of KitNET is the RMSE score, $S$, produced by the aggregate autoencoder. It should be noted that the scores produced by KitNET are numerical values rather than a probability distribution or logits like in standard deep learning classifiers. Kitsune utilizes a classification scheme which triggers an alarm under the condition: $S \geq \phi \beta$, where $\phi$ is the highest value of $S$ recorded during training and $\beta$ is a constant used to find a trade-off between the number of false positives and negatives. The authors limit the value of $\beta$ to be greater than or equal to $1.0$ to assure a $100\%$ training accuracy (i.e., all the training data are considered benign).

%% file: tex/5-setup.tex
\section{Experimental Setup} \label{sec:setup}
In this section, we briefly describe our experimental setup and the necessary modifications to the KitNET. 

\subsection{Implementing KitNET}
In order to perform adversarial machine learning, the original C++ version of Kitsune was reproduced in TensorFlow~\cite{tensorflow}. The TensorFlow model was tested and evaluated similarly to the C++ implementation with an average deviation on the outputs of $5.71 \times 10^{-7}$ from the original model. We then utilized the Cleverhans~\cite{cleverhans}, an adversarial machine learning library that is produced and maintained by domain experts, to mount different adversarial example generation algorithms on the Kitsune. We also used the same Mirai dataset as in~\cite{mirsky2018kitsune}.

\subsection{Modifications to the Model}
Our implementation of KitNET moves the classification mechanism into the model by adding a final layer at the output, as expressed in Equation \ref{eq:classlayer}.
\begin{equation}\label{eq:classlayer}
     C( \vec{x}) = \begin{bmatrix} \mbox{benign} \\ \mbox{malicious}\end{bmatrix} = S(\vec{x}) \begin{bmatrix}1 \\ -1\end{bmatrix} + \begin{bmatrix}0 \\ 2T\end{bmatrix}
\end{equation}
This allows the deep learning model to produce the classification result based on a threshold, $T$. Effectively, this alteration moves the original classification scheme into KitNET itself when $T = \phi \beta$, transforming the model from a regression model into a classifier.

As adversarial examples target deep learning models, we isolate KitNET from Kitsune when performing our attacks. In a real-world attack on Kitsune, the adversary must circumvent, or surmount, the Feature Extractor to induce perturbations on KitNET's input. However, understanding the Feature Extractor makes it feasible for the adversary to craft network traffics to generate essential features. Thus, in our experiments, we focus on evaluating the security of KitNET from the normalized feature space.

%% file: tex/6-networksecurity.tex
\section{Evaluation from the Network Security Perspective}\label{sec:NS}
A DL-NIDS must be evaluated from both the network security and adversarial machine learning aspects to understand its defensive capabilities fully. In the domain of intrusion detection, the ability to distinguish malicious network traffics from benign traffics is the primary performance metric. In this section, we evaluate the classification accuracy of the Kitsune.

Kitsune's developers evaluate the DL-NIDS against a series of attacks in a variety of networks~\cite{mirsky2018kitsune}. In our implementation, the accuracy of Kitsune is highly dependent on the threshold, $T$. This value defines the decision boundary, which makes it a critical parameter when deploying the model. We evaluate the KitNET by assuming that the threshold is not predefined but trained as an end-to-end deep learning system. In addition, this analysis also indicates how the threshold correlates with the perturbation required in adversarial machine learning.

To assess the performance of a given threshold value, we consider the following two metrics:
\begin{enumerate}
  \item \textbf{False Positives:} The percentage of benign inputs that are incorrectly classified as malicious. 
  \item \textbf{False Negatives:} The percentage of malicious data that are incorrectly classified as benign. 
\end{enumerate}
On the one hand, the rate of false positives accounts for the reliability of a network. On the other hand, the rate of false negatives is closely associated with the intrusion detection system's effectiveness. Therefore, both rates should be minimized in an ideal situation. However, in the setting of Kitsune, the value of $T$ acts as a trade-off between false positive rate and false-negative rate.

We investigated the whole functional range of possible thresholds in this analysis, i.e., from the minimum score of $0$ to $20$, which leads to $100\%$ false negatives on the given dataset. Figure \ref{fig:threshold}(a) plots the two metrics as well as the accuracy of the DL-NIDS.

\input{tex/tab-thresholds}
It can be seen that the rates of false positives and false negatives remain almost unchanged in the middle range. Furthermore, it can also be observed that if we want to minimize one of the rates, the other rate will increase significantly. Finally, the accuracy is also essentially unchanged for threshold values below 7, which can partially contribute to the imbalance of the dataset (i.e., most of the data belong to the benign class). Therefore, a threshold between 0.05 and 1 would be appropriate for this scheme. The effectiveness of Kitsune at separating the Mirai dataset is further demonstrated by the ROC curve in Figure \ref{fig:threshold}(b).

%% file: tex/tab-thresholds.tex


\begin{figure}[htbp]
\resizebox{0.49\textwidth}{!}{%
\includegraphics[]{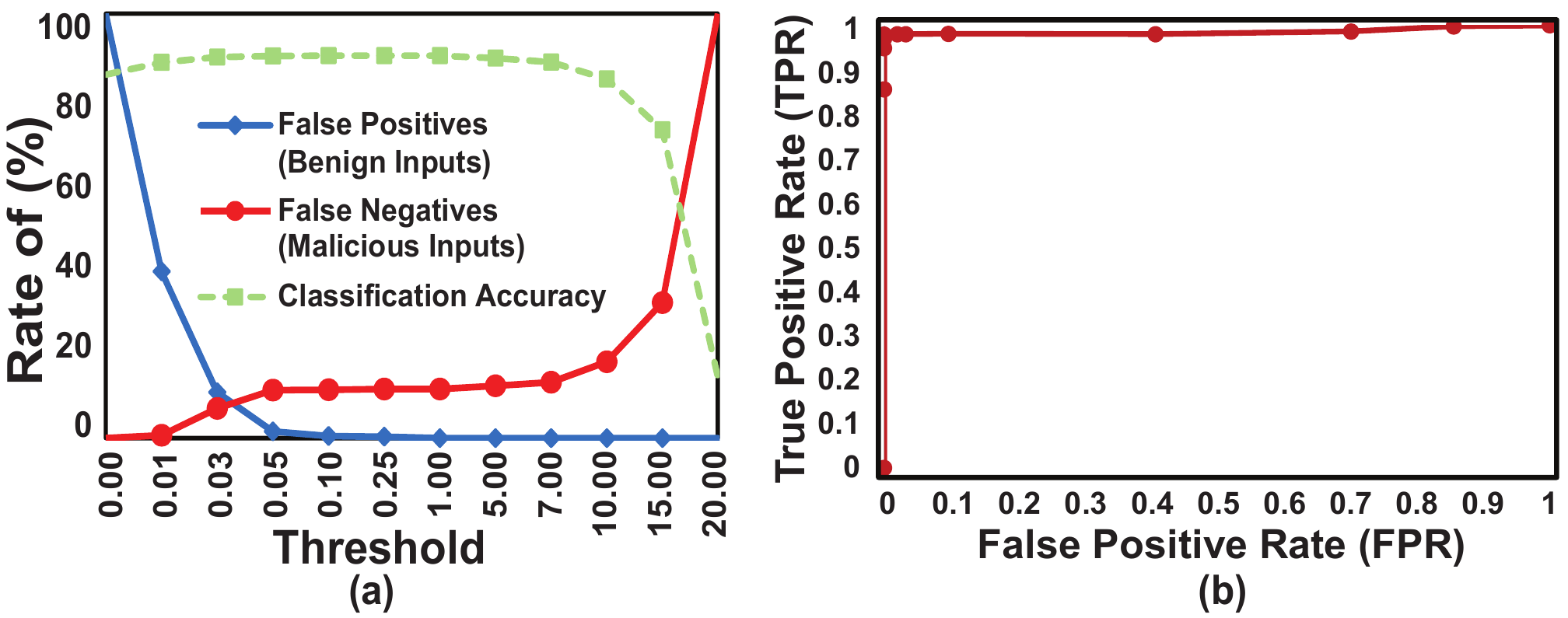}}
\centering
\caption{The percentage of misclassified benign and malicious inputs for chosen threshold values (a). A receiver operating characteristic (ROC) curve for Kitsune (b).}
\label{fig:threshold}
\end{figure}

%% file: tex/7-adversarialML.tex
\section{Evaluation against Adversarial Machine Learning}\label{sec:AM}
This section continues the evaluation of Kitsune through an empirical analysis of its robustness against adversarial examples.

\subsection{Adversarial Example Generation Methods}
Intelligent and adaptive adversaries will exploit the vulnerability of the machine learning models against novel DL-NIDS by using techniques such as adversarial examples and poisoning attacks. There are mainly two attacking objectives in adversarial machine learning, namely, integrity and availability violations. In this setting, integrity violations attempt to generate malicious traffic, which evades detection (produce a false negative), while availability violations attempt to make benign traffic appear malicious (produce a false positive)~\cite{ apruzzese2018effectiveness }. However, adversarial examples attempt to achieve a misclassification with perturbations as small as possible.

Another concern in performing these attacks is that the network data are fundamentally distinct from images, usually used in conventional adversarial machine learning. An adversarial example in the image domain is an image perceived to be the same by human observers but differently by the model. The $L_P$ norm between the two images exemplifies visual distance and can be used as the distance metric. However, this definition fails in network security as observing network traffic at the bit-level is not generally practical. Therefore, the semantic understanding of these attacks in this setting is remarkably different.  

One potential definition for adversarial examples in this scenario, which is facilitated by the architecture of Kitsune, is to use the extracted features generated by the model as an indication of the observable difference. Thus, we adopt the $L_P$ distance on the feature space between the original input and the perturbed input as the distance metric. In particular, the $L_0$ norm correlates to altering a small number of extracted features, which might be a better metric than other $L_P$ norms.

Many methods of generating adversarial examples have been developed. With each thrives in different settings, we attempt to generate a broad comparison of adversarial examples with different distance metrics in the network security domain. We evaluate the robustness of the KitNET against the following algorithms:
\begin{itemize}
  \item \textbf{Fast Gradient Sign Method (FGSM):}
  This method optimizes over the $L_{\infty}$ norm (i.e., reduces the maximum perturbation on any input feature) by taking a single step to each element of $\vec{x}$ in the direction opposite the gradient ~\cite{goodfellow2014explaining}.
  \item \textbf{Jacobian Base Saliency Map (JSMA):}
  This attack minimizes the $L_0$ norm by iteratively calculating a saliency map and then perturbing the feature that will have the highest effect~\cite{papernot2016limitations}.
  \item \textbf{Carlini and Wagner (C\&W):}
  Carlini and Wagner's adversarial framework, as discussed earlier, can either minimize the $L_2$, $L_0$ or $L_{\infty}$ distance metric~\cite{carlini2017towards}. Our experiments utilize the $L_2$ norm to reduce the Euclidean distance between the vectors through an iterative method.
  \item \textbf{Elastic Net Method (ENM):}
  Elastic net attacks are novel algorithms that limit the total absolute perturbation across the input space, i.e., the $L_1$ norm. ENM produces the adversarial examples by expanding an iterative $L_2$ attack with an $L_1$ regularizer~\cite{chen2018ead}.
\end{itemize}

\subsection{Experimental Results}
We conduct our experiments on both integrity and availability violations. Integrity violation attacks are performed on the benign inputs with a threshold of $s=1.0$. The experimental results are presented in Table \ref{tab:integ}. For comparison between different algorithms, the common $L_P$ distance metrics are all presented. Each attack was conducted on the same $1000$ random benign samples from the dataset.

\input{tex/tab-intergirty}

Availability attacks are also performed using the same threshold of $s=1.0$. $1000$ input vectors that yield the closest output scores to the threshold were selected. The results are summarized in Table \ref{tab:avail}. As the normalizers were only trained on benign inputs, many malicious inputs would be normalized outside the typical range between $0$ and $1$. 

\input{tex/tab-availability}

\subsection{Analysis and Discussion}
By comparing Table \ref{tab:integ} and Table \ref{tab:avail}, it can be seen that the integrity attacks, in general, perform much better than the availability attacks. For instance, adversarial examples are rarely generated in the FGSM and JSMA availability attacks. Additionally, the perturbations produced by the availability attacks are all larger than their integrity counterparts. A potential cause for the difficulty is the disjoint nature between the benign and malicious input data, as exhibited by the clipping of the normalized inputs, in conjunction with a boundary decision (i.e., the threshold $T$) that is much closer to the benign input data. 

Among these four methods, the earlier algorithms, i.e., the FGSM and JSMA, perform worse than the C\&W and ENM attacks. As we mentioned above, especially in the availability attacks, the success rates of these attacks are significantly low. This result is expected since the more advanced iterative C\&W and ENM algorithms can search a larger adversarial space than the FGSM and JSMA.

A final observation is that ENM is very effective in these attacks. Even though this attack is optimized for the $L_1$ norm, its generated adversarial examples simultaneously yield minimal values for the other norms. Specifically, the $L_0$ perturbations produced were even better than those produced by JSMA. As stated above, the $L_0$ norm seems to be the most appropriate norm among these four $L_p$ norms in the setting of network security, as it signifies altering a minimized number of extracted features from the network traffic. Thus, ENM can be implemented against the Kitsune to generate adversarial examples to fool the detection system while requiring minimal perturbations.

We note that the above attacks were produced with an adaptive step size random search of the parameters of each method. In practice, adversaries may use such a naive approach to determine effect attack algorithms. Then, utilize more robust optimization algorithms, such as Bayesian or gradient descent optimization, with the indicated attack algorithms to produce a superior result.

%% file: tex/tab-intergirty.tex
\begin{table}[htbp]
\centering
\caption{Integrity Attacks on KitNET}
\label{tab:integ}
\begin{small}
\begin{tabular}{|c|c|c|c|c|c|}
        \hline
        & & \multicolumn{4}{c|}{$L_P$ Distances} \\
        \cline{3-6}
        \hline
        Algorithm & Success (\%) & $L_0$  & $L_1$  & $L_2$  & $L_\infty$   \\
        \hline
        \hline
        FGSM & $100$  & $100$ & $108$ & $10.8$  & $1.8$   \\
        \hline
        JSMA & $100$  & $2.33$ & $10.73$ & $6.97$  & $4.87$   \\
        \hline
        C\&W & $100$  & $100$ & $7.44$ & $3.61$  & $3.49$   \\
        \hline
        ENM & $100$  & $1.21$ & $4.94$ & $4.64$  & $4.49$   \\
        \hline
\end{tabular}
\end{small}
\end{table} 

%% file: tex/tab-availability.tex
\begin{table}[htbp]
\centering
\caption{Availability Attacks on KitNET}
\label{tab:avail}
\begin{small}
\begin{tabular}{|c|c|c|c|c|c|}
        \hline
        & & \multicolumn{4}{c|}{$L_P$ Distances} \\
        \cline{3-6}
        \hline
        Algorithm & Success (\%) & $L_0$  & $L_1$  & $L_2$  & $L_\infty$   \\
        \hline
        \hline
        FGSM & $4$  & $100$ & $78.00$ & $7.79$  & $0.78$   \\
        \hline
        JSMA & $0$  & $-$ & $-$ & $-$  & $-$   \\
        \hline
        C\&W & $100$  & $100$ & $22.00$ & $8.50$  & $5.61$   \\
        \hline
        ENM & $100$  & $8.74$ & $21.7123$ & $8.14$  & $3.60$    \\
        \hline
\end{tabular}
\end{small}
\end{table} 

%% file: tex/8-optimize.tex
\subsection{Optimizing ENM}
Since ENM has been demonstrated to be very successful in our experiments, we next focus on optimizing the ENM attack on Kitsune in our setting. The CleverHans implementation uses a simple gradient descent optimizer to minimize the function:
\begin{equation}\label{eq:enm}
  c \cdot \mbox{max}\{F(\vec{x})_j-Y \mbox{ , } 0\} + \beta ||\vec{x}-\vec{x_0}||_1 +||\vec{x}-\vec{x_0}||_2
\end{equation}
where $F(\cdot)_j$ is the logit output of the target classifier, Y is the target logit output (i.e., the output which produces the desired violation), and $\vec{x_0}$ is the original network input. It can be seen that there are two regularization parameters, $c$ and $\beta$. These parameters determine the contribution of the different metrics to the attack algorithm. For example, a very large $c$ effectively increases the attack's ability to converge to a successful attack. The large contribution of the constraint terms also potentially overshadows the distance metrics, effectively diminishing the attack's ability to minimize the perturbation. The focus of this optimization is to determine optimal regularization terms to produce effective attacks on KitNET.

The ENM algorithm has several other hyper-parameters, including the learning rate, maximum gradient descent steps, and targeted confidence level. These parameters are standard in adversarial example attacks; these parameters are set to the constant values of $0.05$, $1000$, and $0$, respectively. An optimization scheme included in the ENM algorithm aids in producing optimal results by altering $c$. It decreases the parameter $N$-times, only retaining the successful attack, which produces the lowest perturbation. This feature is disabled by setting $N=0$, ensuring that it does not alter optimization results. Therefore, the results of the optimization could be further improved by enabling this functionality.

\begin{figure}[htbp]
\resizebox{0.49\textwidth}{!}{%
\includegraphics[]{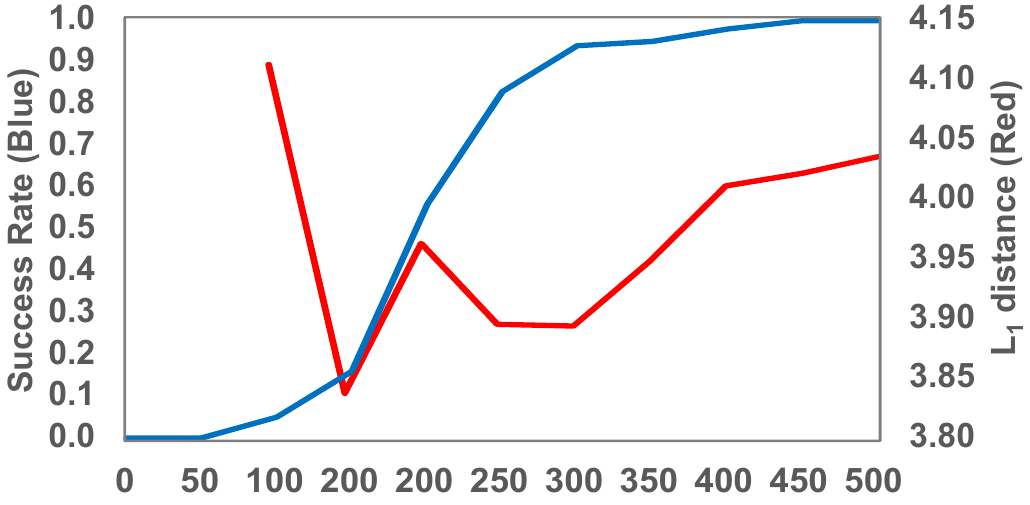}}
\centering
\caption{The success rate (blue) and average $L_1$-distance (red) of adversarial examples with respect to the regularization parameter, $c$, used for the attack.}
\label{fig:C}
\end{figure}

The parameter, $c$, determines the contribution of the adversarial misclassification objective at the cost of diminishing the two $L_P$ normalization terms. Thus, it can be logically determined that the optimal value of $c$ is that value that achieves the demanded success rate while remaining as small as possible. We evaluate a wide range of $c$ values for $\beta=1$, as shown in Figure \ref{fig:C}. We find $c=450$ optimal, which achieves a $100\%$ success rate with a relatively small perturbation. It can also be observed from Figure \ref{fig:C} that the resultant $L_1$ distance does not directly correlate to the selection of $c$. We also tried to increase the value of $c$ into the thousands; interestingly, the $L_P$ distances still only changed very slightly. 

On the other hand, the choice of $\beta$ significantly affects the $L_P$ distances. We now optimize the produced perturbation through varying the parameter $\beta$ for $c=450$. The results are summarized in Table \ref{tab:beta}. It can be seen that the success rate will drop as the increase of $c$, after the second term of Equation \ref{eq:enm} begins to overpower the loss function associated with $c$.

\input{tex/tab-beta.tex}
 
\vspace{1em}

\textbf{Summary:} It can be concluded that adversarial machine learning can be a real threat against DL-NIDS. Therefore, when moving intrusion detection towards the profound learning realm, it is critical to evaluate the security of a DL-NIDS against both adversarial attacks in the conventional network and the machine learning domains. 

%% file: tex/tab-beta.tex
\begin{table}[htbp]
\centering
\caption{The perturbations produced with respect to $\beta$.}
\label{tab:beta}
\begin{small}
\begin{tabular}{|c|c|c|c|c|c|}
        \hline
        & & \multicolumn{4}{c|}{$L_P$ Distances} \\
        \cline{3-6}
        \hline
        $\beta$ & Success (\%) & $L_0$  & $L_1$  & $L_2$  & $L_\infty$   \\
        \hline
        \hline
        $1 \times 10^{-5}$ & $100$  & $96.61$ & $5.9518$ & $3.6378$  & $3.5163$   \\
        \hline
        $1 \times 10^{-4}$ & $100$  & $78.46$ & $5.7574$ & $3.6388$  & $3.516$   \\
        \hline
        $1 \times 10^{-3}$ & $100$  & $33.34$ & $5.0577$ & $3.6435$  & $3.5268$   \\
        \hline
        $1 \times 10^{-2}$ & $100$  & $5.51$ & $5.1722$ & $3.7658$  & $3.3129$   \\
        \hline
        $1 \times 10^{-1}$ & $100$  & $1.09$ & $3.8624$ & $3.7277$  & $3.6450$   \\
        \hline
        $1 \times 10^0$ & $100$  & $1.01$ & $4.0347$ & $4.0158$  & $4.0044$   \\
        \hline
        $2 \times 10^1$ & $0.84$  & $1.00$ & $4.1350$ & $4.1350$  & $4.1350$   \\
        \hline
        $5 \times 10^1$ & $0.08$  & $1.00$ & $4.2054$ & $4.2054$  & $4.2054$   \\
        \hline
        $1 \times 10^2$ & $0$  & - & - & - & -   \\
        \hline
\end{tabular}
\end{small}
\end{table} 

%% file: tex/999-conclusions.tex
\section{Conclusions and Future Directions} \label{sec:conclusions}
This paper has demonstrated the vulnerability of DL-NIDS to well-crafted attacks from the domain of adversarial machine learning. This vulnerability is present in deep learning-based systems even when the model achieves high accuracy for classifying benign and malicious network traffic. Therefore, researchers must take steps to verify the security of deep learning models in security-critical applications to ensure they do not impose additional risks; otherwise, it will defeat the purpose of using deep learning techniques to protect networks. 

The existence of the Feature Extractor and the Packet Parser signifies that the Kitsune is at least partially utilizing domain knowledge of network traffic to generate its classification. Their applications strive to be as data-driven as possible to get the most benefit from deep learning models (i.e., they require little to no human knowledge to generate a function mapping). Thus, despite the current success of Kitsune and other DL-NIDS, as the field continues to develop, DL-NIDS will attempt directly converting network traffic to a classification utilizing end-to-end deep learning models. Furthermore, the human knowledge currently being used by modern DL-NIDS implies that to increase the probability of a successful attack, an adversary should understand this knowledge. Thus, as DL-NIDS continues to develop, evaluating the model against adversarial machine learning techniques becomes even more critical as attacks will no longer require this additional knowledge when targeting the system.

This work assumes that the adversary has direct knowledge of the target DL-NIDS, allowing them to directly generate inputs for the deep learning model. A potential drawback of this assumption is that the perturbation requires to generate the adversarial examples does not directly correlate to the alteration on the network. Additionally, it does not account for the effect of that change on the network traffic on the host device. Future works will address this gap between the adversarial input to the deep learning model and the network traffic.

%% file: tex/999-acknowledgements.tex
\section*{Acknowledgement}
This work is partially supported by the National Science Foundation award 2047384.